# Leaky Integrate-and-Fire Neuron under Poisson Stimulation.

Kseniia Kravchuk

Synergetics department
Bogolyubov institute for Theoretical physics
Kyiv, Ukraine
kgkravchuk@bitp.kiev.ua

*Abstract* —We consider a single Leaky integrate-and-fire neuron stimulated with Poisson process. We develop a method, which allows one to obtain the first passage time probability density function without any additional approximations.

I. INTRODUCTION

If the input stimulation of a neuron were known, what would be the output? This problem has been addressed many times during the last 5 decades, see e.g. [1]–[3]. A widely used way to deal with this problem is to apply the diffusion approximation. In the diffusion approximation, it is assumed, that each input impulse, received by a neuron, has an infinitesimally small effect on the neuron's state (on its membrane voltage). If so, the number of input impulses needed to trigger the neuron is infinite. Nevertheless, a careful consideration of experimental data shows, that the number of input impulses, needed to trigger a real neuron, is not always sufficiently large to support the diffusion approximation [4]–[6]. Moreover, the diffusion approximation does not allow a consistent theoretical consideration of a neural network, when the output impulses of some neuron are sent to the input of another neuron.

That is why, one should cast out the diffusion approximation and consider a more realistic situation, when each input impulse causes a significant change in neuron's membrane voltage, and the number of input impulses, needed to trigger the neuron, is finite. Recently, the first steps were made in this direction for the Binding neuron model [7] and for the Leaky integrate-and-fire (LIF) neuron model [8] stimulated with Poisson stream of input impulses. In [8], the first passage time probability density function (p.d.f.) is derived for the particular case, when 2 impulses are enough to trigger the neuron. Unfortunately, the approach developed in [8] is too complicated and cannot be extended to the general case. In this paper, we develop a general method for the LIF model, which allows to obtain exact analytical results for the first passage time p.d.f. without any additional approximations.

II. PROBLEM STATEMENT

Consider a separate neuron, stimulated by a sequence of input impulses. Intervals between input impulses are stochastic variables. As the probability distribution for the input interspike intervals we take Poisson p.d.f. with intensity λ. Assume that any input impulse has a significant effect on neuron's state, and that the amount of impulses needed to trigger the neuron is finite. The neuron is modelled with the LIF model. We find rigorously a probability density $P(t)$ for the first passage time of this system. Namely, we assume that the neuron starts from its resting state at 0 moment of time, and find the probability $P(t)dt$ that the threshold will be reached for the first time at the moment *t*, with precision *dt* (*t* is called the first passage time).

A. *Leaky Integrate-and-Fire neuronal model*

At any moment of time, the state of LIF neuron is completely determined by a single real variable *V*, which resembles the membrane voltage of a real neuron. The resting state is associated with $V=0$. Each input impulse raises the membrane voltage by a fixed quantity *h*:

$$V(t_i^+) = V(t_i^-) + h, \quad (1)$$

were $t_i$ denotes an arrival time of *i*-th input impulse, *i*=1,2,.... Between input impulses, the membrane voltage decays exponentially with a character time τ:

$$V(t_i + \Delta t) = V(t_i^+)\, e^{-\Delta t/\tau}, \quad (2)$$

were Δ*t* is a time interval free of input impulses.

LIF neuron fires an output spike, when the membrane voltage reaches its threshold value $V_0$, and then the neuron comes back to its resting state $V=0$.

III. THE METHOD

A. *Zones of membrane voltage values*

Depending on its parameters, LIF neuron may require different number of input impulses to reach the threshold and fire an output spike. Let us denote with *N* the minimal number of input impulses, needed to trigger the neuron. An expression for *N* can be easily derived from neuron parameters:

$$N = [V_0/h] + 1, \quad (3)$$

where the square brackets denote an integer part of the corresponding expression. In particular realizations of an input stochastic process, the number of input impulses, which trigger the neuron, may be different, depending on impulses' arrival times, but never less than *N*.

Let us divide the range of membrane voltage values, $V \in [0; V_0]$, into zones of height *h*, beginning from the top, see

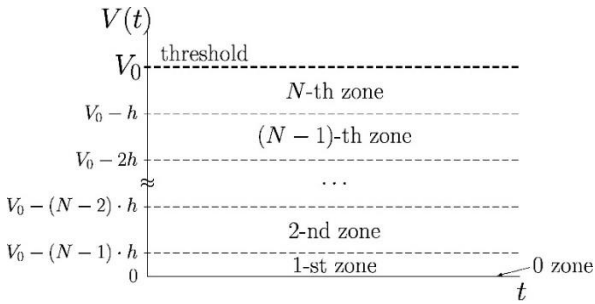

Fig. 1. Zones of membrane voltage values.

Fig.1. We obtain ($N$–1) zones of height $h$ and one more zone of a smaller height at the bottom. The zones can be assigned with numbers, e.g. as at the Fig. 1:

1-st zone:  $V \in (0; V_0 - (N-1) \cdot h)$,

$i$-th zone:  $V \in [V_0 - (N-i+1) \cdot h; V_0 - (N-i) \cdot h)$,    $i = 2, 3, \ldots, N$.  (4)

The resting state, $V=0$, should be considered as a separate zone, because of nonzero probability, associated with it. So, let us define a 0 zone as follows:

0 zone:    $V = 0$.    (5)

Let us introduce probabilities $p(i,t)$, $i = 0,1,\ldots,N$ to have the membrane voltage within $i$-th zone at the moment $t$, given the neuron didn't reach the threshold before $t$. We would like to emphasize, that we are not going to find probabilities of the specific values of $V$ at different moments of time – just the probabilities to have the membrane voltage within the certain limits, see (4). Such simplification can be made without any loose of precision in $P(t)$. Indeed, let us imagine, that at some moment of time the membrane voltage takes its value within $N$-th zone: $V \in [V_0 - h; V_0)$. If the neuron obtains an input impulse at this moment, then the membrane voltage will reach its threshold, and an output spike will be generated. The generation of an output impulse will not depend on the exact value of $V$, since $V$ belongs to the $N$-th zone. Therefore, output ISI probability density can be found as a product of $p(N,t)$ and the probability to obtain an input impulse within time interval $[t;t+dt)$, which is $\lambda dt$ for a Poisson process:

$P(t)dt = p(N,t) \cdot \lambda dt$,    (6)

were $\lambda$ is the intensity of input Poisson process. So, the first passage time problem reduces to finding $p(N,t)$.

*B. Equations*

It seems, that the probability $p(N,t)$ should be found together with probabilities $p(i,t)$ for the rest of zones, because the membrane voltage may get across the zone's borders. The mechanisms of such crossings are

- jumps,
- exponential decay.

The first one is activated with the arrival of input impulse, and guarantees, that $V$ will move for one zone up. The second one is activated permanently, and provides the possibility to move for one zone down. In order to account the second mechanism, let us introduce $l_{i,i-1}(t)$ as a flow of probability across the zones' borders due to the exponential decay. Namely, $l_{i,i-1}(t)\,dt$ gives the probability that at the moment $t$ with precision $dt$ the trajectory of membrane voltage will cross the border of $i$-th and $(i-1)$-th zones due to the exponential decay. In calculation of $l_{i,i-1}(t)$, only trajectories which have never reached the threshold should be accounted.

Now, all the transitions between zones can be written as the following system of ODEs:

$$\begin{cases} \dfrac{dp(0,t)}{dt} = -\lambda p(0,t), \\ \dfrac{dp(1,t)}{dt} = -\lambda p(1,t) + l_{2,1}(t), \\ \dfrac{dp(2,t)}{dt} = -\lambda p(2,t) + \lambda p(1,t) + \lambda p(0,t) - l_{2,1}(t) + l_{3,2}(t) \\ \dfrac{dp(i,t)}{dt} = -\lambda p(i,t) + \lambda p(i-1,t) - l_{i,i-1}(t) + l_{i+1,i}(t), \\ \qquad\qquad\qquad\qquad\qquad\qquad i = 3, \ldots, N-1, \\ \dfrac{dp(N,t)}{dt} = -\lambda p(N,t) + \lambda p(N-1,t) - l_{N,N-1}(t). \end{cases}$$    (7)

Here, the terms like $\lambda p(i,t)$ in the right-hand side correspond to the jumping out of the $i$-th zone upward with the arrival of input impulse (see the first mechanism, above). The terms like $l_{i,i-1}(t)$ describe the probability flow across the zones' borders due to the exponential leak. This system should be supplemented with initial conditions:

$p(0,t)|_{t=0} = 1$,

$p(i,t)|_{t=0} = 0$,    $i = 1, 2, \ldots, N$.    (8)

In order to solve (7) and to obtain $p(N,t)$, one should first find expressions for all $l_{i,i-1}(t)$.

*C. The flow of probability across the zones borders*

The crossing of a given zones' border at definite time $t$ may happen in different trajectories. Such trajectories may contain a different number $k$ of input impulses, obtained within time interval $(0;t)$, and the arrival times of such impulses may differ as well. That is why, it is natural to find $l_{i,i-1}(t)$ as a sum by $k$:

$l_{i,i-1}(t) = \sum_{(k_{min})_i}^{(k_{max})_i} l^k_{i,i-1}(t)$    (9)

where functions $l^k_{i,i-1}(t)$ correspond to trajectories with $k$ input impulses within interval $(0;t)$. Namely, $l^k_{i,i-1}(t)\,dt$ gives the probability of crossing the border of $i$-th and $(i-1)$-th zones at time $t$ with precision $dt$ due to exponential decay (only trajectories which never reached the threshold should be accounted).

*D. Minimum and maximum number of input impulses*

The number $k$ of input impulses which contribute to $l_{i,i-1}(t)$ in (9) should be limited both from the top and from the bottom. Indeed, $k$ should be sufficiently large to let the membrane voltage reach the zones' border (which in accordance with (4) is determined by equality $V = V_0 - (N - i + 1) \cdot h$):

$(k_{min})_i = [(V_0 - (N - i + 1) \cdot h) / h] + 1 = i - 1$,    (10)

where the definition (3) is taken into account.

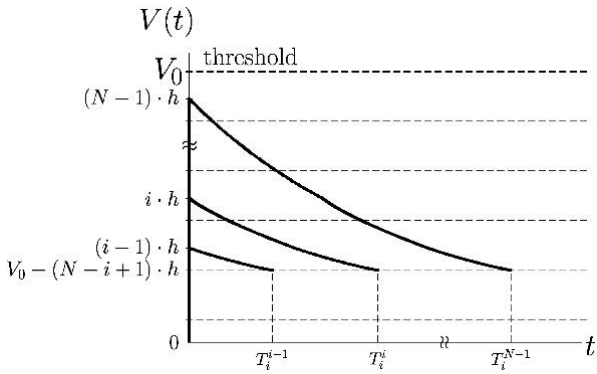

Fig. 2. $T_i^j$ is a time interval needed to the membrane voltage to decay from the value $V=j \cdot h$ to the border of $i$-th and $(i-1)$-th zones, $V=V_0-(N-i+1) \cdot h$.

On the other hand, $k$ cannot be arbitrarily large, because otherwise the threshold will be reached before the moment $t$. This imposes an upper limit for the possible values of $k$. In order to find such limit, let us first define two sets of time intervals with the following expressions:

$$T_i^j \equiv \tau \ln \frac{j \cdot h}{V_0 - (N-i+1) \cdot h}, \quad j = i-1, i, \ldots, N-1,$$

$$\tilde{T}_i \equiv \tau \ln \frac{V_0}{V_0 - (N-i+1) \cdot h}, \quad i = 2, 3, \ldots, N. \quad (11)$$

$T_i^j$ is a time interval, needed for the membrane voltage to decay from the level $V=j \cdot h$ to the border of $i$-th and $(i-1)$-th zones (see Fig. 2), and $\tilde{T}_i$ is the time interval which it takes to the membrane voltage to decay from $V_0$ to the same border.

Now, consider the closest packing of input impulses within time $t$, which does not cause triggering and allows crossing the zones border afterwards, see Fig.3. The first $(N-1)$ impulses may arrive arbitrarily close in time to one another, and they would not be able to trigger the neuron. And the others should be separated with time intervals, needed to the membrane voltage to decay to the level $V_0-h$, to avoid triggering. After the $(k_{max})_i$-th impulse there should be a time interval $\tilde{T}_i$ to let the membrane voltage drop from $V=V_0^-$ to the level $V=V_0-(N-i+1)\cdot h$ of zones' border, see Fig. 3.

Therefore, one obtains $(k_{max})_i = N + \left[\frac{t - T_{N-1}^N - \tilde{T}_i}{\tilde{T}_N}\right]$, which can be rewritten using (11) as follows:

$$(k_{max})_i = N - 1 + \left[\frac{t - T_i^{N-1}}{\tilde{T}_N}\right], \quad t \geq T_i^{N-1} + \tilde{T}_N. \quad (12)$$

Similarly, if $t < T_i^{N-1} + \tilde{T}_N$, $(k_{max})_i$ can be obtained as

$$(k_{max})_i = i - 1 + j, \quad t \in [T_i^{i-1+j}; T_i^{i+j})_j, \quad j = 0, 1, \ldots, N-i-1 \quad (13)$$

Finally, we have obtained, that $(k_{max})_i$ increases with the increase in $t$ and changes its value by unity at the points

$$T_i^{i-1}, T_i^i, \ldots, T_i^{N-1}, T_i^{N-1} + \tilde{T}_N, T_i^{N-1} + 2\tilde{T}_N, T_i^{N-1} + 3\tilde{T}_N, \ldots$$

and the corresponding values of $(k_{max})_i$ are $i-1, i, \ldots, N-1, N, N+1, N+2, N+3, \ldots$

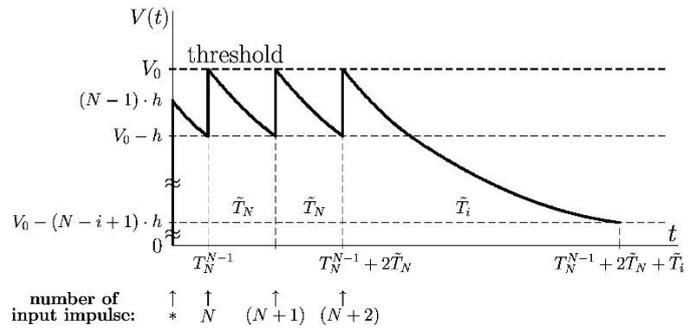

Fig. 3. The closest packing of $(N+2)$ input impulses, which does not cause triggering. * – the first $(N-1)$ impulses may arrive arbitrarily close to zero time moment.

If $t < T_i^{i-1}$, the probability flow across the zones' border is zero:

$$l_{i,i-1}(t) = 0, \quad t < T_i^{i-1}, \quad (14)$$

because any trajectory will not have enough time to reach the border. If $t \geq T_i^{i-1}$, the trajectories may manage to cross the zones' border and the probability flow emerges. Therefore, in this case, one should obtain all the $l^k_{i,i-1}(t)$ in order to find $l_{i,i-1}(t)$.

E. Functions $l^k_{i,i-1}(t)$

Trajectories contribute to $l^k_{i,i-1}(t)$ if the following conditions are satisfied:

1. time interval $(0;t)$ contains $k$ input impulses;
2. trajectory crosses the border between $i$-th and $(i-1)$-th zones at time $\underline{t}$ with precision $dt$;
3. trajectory haven't reached the threshold until the moment $t$.

Then $l^k_{i,i-1}(t)$ will be a measure of trajectories which satisfy conditions (1)–(3), above. This measure can be found as follows.

Let us denote with $t_1, t_2, \ldots, t_k$ the arrival times of input impulses in a trajectory which contributes to $l^k_{i,i-1}(t)$. The probability to obtain from the Poisson stream $k$ input impulses within time interval $(0;t)$ at the moments $t_1, t_2, \ldots, t_k$ equals $\lambda^k e^{-\lambda t}$. In order to obtain $l^k_{i,i-1}(t)$ one should integrate the this expression over all possible values of arrival times $t_j$:

$$l^k_{i,i-1}(t) = \int_{\Omega_T} \ldots \int dt_2 \ldots dt_k \; \lambda^k e^{-\lambda t}, \quad (15)$$

where the integration domain $\Omega_T$ is defined by conditions 2 and 3, above. In (15), the integration over $t_1$ is missing, because the variables $t_1, t_2, \ldots, t_k$ are not independent. Indeed, condition 2 ensures that one of the arrival times can be found using all the others, e.g. one can find $t_1$ as a function of $t_2, \ldots, t_k$ and $t$.

Domain $\Omega_T$ in the space with coordinates $t_2, \ldots, t_k$ has a complicated structure, due to the conditions 2 and 3, above. This problem may be simplified by appropriate variable change. Let us introduce variables $V_2, V_3, \ldots, V_k$, where $V_j$, $j=2,\ldots,k$, denotes the membrane voltage just after the arrival of $j$-th input impulse:

$$V_j \equiv V(t_j^+), \quad j = 2, 3, \ldots, k. \quad (16)$$

The mapping $t_2,\ldots,t_k \to V_2,\ldots,V_k$ is one-to-one, so we can perform the corresponding variable change in (15):

$$l_{i,i-1}^k(t) = \lambda^k e^{-\lambda t} \int \ldots \int_{\Omega_V} \left|\frac{\partial(t_2,\ldots,t_k)}{\partial(V_2,\ldots,V_k)}\right| dV_2 \ldots dV_k, \quad (17)$$

where the integration domain $\Omega_V$ in a space with coordinates $V_2,\ldots, V_k$ corresponds to the domain $\Omega_T$ in a space with coordinates $t_2,\ldots, t_k$ and is defined by the conditions 2 and 3, above. The Jacobean in (17), can be easily found using (1), (2) and (16):

$$\left|\frac{\partial(t_2,\ldots,t_k)}{\partial(V_2,\ldots,V_k)}\right| = (-\tau)^k \cdot \prod_{j=2}^k \frac{1}{V_j - h}. \quad (18)$$

Such variable change makes it possible to find upper limits of integration easily. Indeed, the first $N-1$ impulses may be located arbitrarily close to each other, therefore, $V_j < j \cdot h$ for $j = 2,3,\ldots, N-1$. Then, after the $N$-th input impulse, the threshold can be reached and the condition 3 should be accounted. In terms of the membrane voltage, this condition can be written as the simple inequality $V_j < V_0$, $j = N, N+1,\ldots, k$. Finally, one can write formally:

$$\Omega_V = \{(V_2,\ldots,V_k) | V_j \in [\underline{V_j}, \overline{V_j}]\}, \quad j = 2,3,\ldots,k, \quad (19)$$

were the upper limits are

$$\overline{V_j} = \begin{cases} j \cdot h, & j = 2,3,\ldots, N-1, \\ V_0, & j = N, N+1,\ldots, k. \end{cases} \quad (20)$$

and the lower limits $\underline{V_j}$ are still to be found using the conditions 2 and 3, above.

In order to find $\underline{V_j}$, one should rewrite the condition 2 in terms of membrane voltage variables $V_j$, using (1) and (2):

$$\prod_{j=2}^k \frac{V_{j-1}}{V_j - h} = \frac{V_0 - (N-i+1)\cdot h}{V_k} \cdot e^{\frac{t-t_1}{\tau}}. \quad (21)$$

The lower limit $\underline{V_j}$ is achieved, if all the previous impulses, $m = 1,\ldots, j-1$, are located as early as possible, given the conditions 1–3, above. Let us denote with $\tilde{V}_m$, $m = 1,2,\ldots,j-1$, the membrane voltage after the arrival of $m$-th input impulse in such arrangement. The first $(N-1)$ impulses may be located as close as possible to the moment 0,

$$t_m = 0, \quad m=1,2,\ldots, N-1, \quad (22)$$

and the others should be separated with time intervals, needed to membrane voltage to decay to the level $V_0-h$, as at Fig. 3. That is why,

$$\tilde{V}_m = \begin{cases} m \cdot h, & m = 1,2,\ldots, N-1, \\ V_0^-, & m = N, N+1,\ldots, j-1. \end{cases} \quad (23)$$

Substituting values of $\tilde{V}_m$ from (23) and $t_1=0$ from (22) to (21), one obtains the lower limit $\underline{V_j}$ as a function of $V_{j+1},\ldots,V_k$:

$$\underline{V_j} = \frac{h}{1 - f_{ij}^k(V_{j+1},\ldots,V_k)}, \quad j = 2,3,\ldots, k, \quad (24)$$

were $f_{ij}^k(V_{j+1},\ldots,V_k)$ is defined by the following expressions:

$$f_{ij}^k(V_{j+1},\ldots,V_k) =$$

$$= e^{-\frac{t-T_i^{N-1}-(j-N)\cdot \overline{T}_N}{\tau}} \cdot \frac{V_k}{V_{j+1}-h} \prod_{m=j+2}^k \frac{V_{m-1}}{V_m-h}, \quad k > N, \quad j \geq N,$$

$$= e^{-\frac{t-T_i^j}{\tau}} \cdot \frac{V_k}{V_{j+1}-h} \prod_{m=j+2}^k \frac{V_{m-1}}{V_m-h}, \quad j < N,$$

$$= e^{-\frac{t-T_i^k}{\tau}}, \quad j = k < N,$$

$$= e^{-\frac{t-T_i^{N-1}}{\tau}}, \quad j = k = N,$$

$$= e^{-\frac{t-T_i^{N-1}-(k-N)\cdot \overline{T}_N}{\tau}}, \quad j = k > N. \quad (25)$$

So, finally, in order to obtain the first passage time p.d.f., one should 1) find all the $l_{i,i-1}^k(t)$, using expressions (17)–(20), (24) and (25); 2) for any fixed $i$, sum up all the $l_{i,i-1}^k(t)$ and obtain the probability flow $l_{i,i-1}(t)$ across the zones' border, using eqs. (9), (10), (12) and (13); 3) solve the system (7) of ODEs for the probabilities $p(i,t)$ to have the membrane voltage within $i$-th zone; 4) having $p(N,t)$, obtain expression for the first passage time p.d.f., using (6).

IV. CONCLUSSIONS

In a present paper, we have considered a single leaky integrate-and-fire neuron stimulated with Poisson stream of input impulses. We have developed a method, which allows one to obtain rigorous expressions for the first passage time p.d.f. for such neuron. The expressions obtained will be different for different relations between neuron's threshold and the altitude of input impulse in the LIF model. In a future, we intend to apply the developed method to a particular case of such relation and to obtain specific expressions for the first passage time p.d.f. in this case.